\documentstyle[aps]{revtex}
\tightenlines
\begin{document}
\title{Homodyne Measurements on a Bose-Einstein Condensate}
\author{J. F. Corney and G. J. Milburn}
\address{Department of Physics, The University of Queensland, QLD 4072,
Australia}
\date{\today}
\maketitle
\begin{abstract}
We investigate a non-destructive measurement technique  to monitor
Josephson-like oscillations 
 between two spatially separated neutral atom Bose-Einstein condensates.  One
condensate is
placed in an optical cavity, which is strongly driven by a coherent optical
field.  The cavity
output field  is monitored using a homodyne  detection scheme. The cavity field
is well detuned
from an atomic resonance, and experiences a dispersive phase shift proportional
to the number of
atoms in the cavity.  The detected current is modulated by the coherent
tunneling oscillations of
the condensate.  Even when there is an equal number of atoms in each well
initially, a phase is
established by the measurement process and Josephson-like oscillations develop
due to measurement
back-action noise alone. 
\end{abstract}
\newpage

\section{Introduction}
The experimental observation of Bose-Einstein condensation (BEC) in
dilute systems of trapped neutral atoms
\cite{Anderson95,Hulet95,Ketterle95,BEC97} 
has stimulated a large research program on Bose-Einstein condensation of 
dilute neutral atom gases in confining potentials.  One aspect of BECs that has
attracted much theoretical work is the idea of phase.  Several papers 
\cite{JavYoo96,CirGarZol96,JacColWal96,Burnett96,GraWonColTanWal98} have discussed
the role of measurements in establishing the phase, in the form of interference
between two condensates or Josephson-like coherent tunneling between the 
condensates.  The latter situation is discussed in this paper, where we 
investigate homodyne detection of the
output of an optical cavity containing one condensate in a double well system. 
The measurement process induces tunneling oscillations though back-action noise
and thus induces
a phase difference  between the two separated parts of the condensate.  In a
self-consistent
manner, the tunneling imposes a phase modulation on the light field, which is
detected in the
homodyne current.

\section{Condensate Model}
The model of the condensate used here, namely the BEC in a double well
potential, has been presented in previous papers\cite{MilCorWriWal97}, and so we
only present a brief overview of the system here. The potential is symmetric
with barrier 
height and well separation chosen so that only two single-particle states 
are below the barrier separating the two wells. This enables a treatment 
of the many body problem with a two-mode approximation. The resulting 
model is sufficiently simple to enable an analytic solution to be found 
for the semiclassical equations, and to permit a tractable numerical 
comparison of the semiclassical description with the full quantum 
dynamics. 
 
Consider a dilute gas of atoms moving in the double well potential, 
$V({\bf r})$ with,
\begin{equation}
V({\bf r})=b\left (x^2-\frac{d}{2b}\right )^2 + \frac{1}{2}m\omega_t^2(y^2+z^2) 
,
\end{equation}
where the inter-well coupling occurs along x, and $\omega_t$ is the trap
frequency in the
$y-z$ plane.  This potential has elliptic fixed points at
${\bf r}_1=+q_0{\bf x}, {\bf r}_2=-q_0{\bf x}$,
where $q_0^2=d/2b$, at which the linearised motion is harmonic with frequency
$\omega_0=(4d/m)^{1/2}$.  We set $\omega_t=\omega_0$ for simplicity.
It is convenient to scale the length in units of the position
uncertainty in a harmonic oscillator ground state, $r_0$ where
$r_0=\sqrt{\hbar /2m\omega_0}$.  The barrier height is then given by
$B=(\hbar\omega_0/8)(q_0/r_0)^2$.

The many-body Hamiltonian describing an atomic BEC in a confining potential is
\cite{GriSnoStr95}
\begin{equation}
\hat H(t)=\int d^3{\bf r}\left[\frac{\hbar^2}{2m}\nabla\hat\psi^\dagger
\cdot\nabla\hat\psi+V\hat\psi^\dagger\hat\psi+\frac{U_0}{2}
\hat\psi^\dagger\hat\psi^\dagger\hat\psi\hat\psi \right ]  ,
\label{Hamiltonian}
\end{equation}
where $m$ is the atomic mass, $U_0=4\pi\hbar^2a/m$ measures the strength of the
two-body interaction, and $a$ is the s-wave scattering length,
$\hat\psi({\bf r},t)$ and $\hat\psi^\dagger({\bf r},t)$ are the
Heisenberg picture field operators which annihilate and create atoms
at position ${\bf r}$, and normal ordering has been used.

For a suitable choice of $B$, only two
energy eigenstates lie beneath the barrier which enables a many body 
treatment in terms of only two single-particle states. For details we 
refer to \cite{MilCorWriWal97}.
 We now define the state $u_0({\bf r})$
as the normalised ground-state mode of the local potential $\tilde{V}^{(2)}({\bf
r})$, around the bottom of each well, with 
energy $E_0$, and define the local mode solutions
of the individual wells $u_{1,2}({\bf r})=u_0({\bf r}-{\bf r}_{1,2})$.
These local modes are approximately orthogonal with a first order 
correction ($\epsilon^1$) to orthogonality given by  the 
overlap between the modes of opposite wells.  The energy eigenstates of the 
global double-well potential may then be
approximated as the symmetric (+) and asymmetric (-) combinations
\begin{equation}
u_\pm({\bf r})\approx \frac{1}{\sqrt{2}}\left [u_1({\bf r}) \pm u_2({\bf
r})\right ]  ,
\label{energyestates}
\end{equation}
with corresponding eigenvalues $E_\pm=E_0\pm{\cal R}$, and
\begin{equation}
{\cal R} =\int d^3{\bf r} u_1^*({\bf r})
[V({\bf r})-\tilde{V}^{(2)}({\bf r}-{\bf r}_1)]u_2({\bf r})  .
\end{equation}
The matrix element ${\cal R}$, which is of order $\epsilon^1$, describes the
coupling between the local modes.  The tunneling frequency, $\Omega$, between 
the two minima is then given by the energy level splitting of these two lowest 
states:
\begin{equation}
\Omega=2{\cal R}/\hbar = \omega_0\frac{q_0^2}{2r_0^2}e^{\frac{q_0^2}{2r_0^2}}.
\end{equation}
In the two-mode approximation we expand the field operators in terms of the 
local modes and introduce the Heisenberg picture annihilation and creation 
operators 
\begin{equation}
c_j(t)=\int d^3{\bf r}u_j^*({\bf r})\hat\psi({\bf r},t)
\label{cre_ani}
\end{equation}
so that $[ c_j,c_k^\dagger]\approx\delta_{jk}$. The validity of this expansion
if ensured when the overlap is small:
\begin{equation}
\frac{{\cal R}}{E_0} = \frac{\Omega}{\omega_0} <<1. 
\end{equation}
The ratio of the separation of the minima of the global potential $V({\bf r})$
to the position uncertainty in the state $u_0({\bf r})$ can be as small as
$2q_0/r_0 = 6$ (as in the simulations presented in this paper), and this
condition is still satisfied.  
The many-body Hamiltonian then reduces to the following two-mode approximation:
\begin{eqnarray}
\hat H_2(t)&=&E_0(c_1^\dagger c_1+c_2^\dagger c_2)
+\frac{\hbar\Omega}{2}(c_1c_2^\dagger+c_1^\dagger c_2)
\nonumber \\
&+&\hbar{\kappa}\left 
((c_1^\dagger)^2c_1^2+(c_2^\dagger)^2c_2^2\right )  ,
\label{modalham}
\end{eqnarray}
where $\kappa=U_0/2\hbar V_{eff}$, and
$V_{eff}^{-1}=\int d^3 {\bf r}|u_0({\bf r})|^4$
is the effective mode volume of each well.

The two-mode approximation is valid when many-body interactions
produce only small modifications of the ground state properties
of the individual potentials.  This is true when
\begin{equation}
\hbar\omega_0=\frac{\hbar^2}{2mr_0^2}>>\frac{N|U_0|}{V_{eff}}  .
\end{equation}
Using $V_{eff}\approx 8\pi^{3/2}r_0^3$ for this case, we obtain
the following condition on the number of atoms
\begin{equation}
N<<\frac{r_0}{|a|}  .
\end{equation}
The values used in our simulations, namely $r_0=5$ $\mu$m, $a=5$ nm and
$N= 100$ satisfy this criterion.  Thus the two-mode approximation is valid 
for small number of atoms compared to current experiments with
$N=10^3-10^6$ .

The first term in Eq. \ref{modalham} may be removed by transforming to 
an interaction picture, resulting in 
the Hamiltonian,

\begin{equation} 
H_{I} = \frac{\hbar \Omega}{2} ( c_{1} c_{2}^{\dag} +
c_{1}^{\dag} c_{2} ) + \hbar \kappa ( (c_{1}^{\dag})^{2} c_{1}^{2} +
(c_{2}^{\dag})^{2} c_{2}^{2} ).
\label{twomodeham}
\end{equation}

A full quantum analysis of the quantum dynamics resulting from the many 
body Hamiltonian Eq(\ref{Hamiltonian}) is not tractable, however 
considerable insight can be gained within the two-mode approximation.
In \cite{MilCorWriWal97} an angular momentum model was defined which is 
equivalent to the Hamiltonian Eq(\ref{twomodeham}). Using the 
transformations
\begin{eqnarray}
\hat{J}_z & = &   \frac{1}{2}(c_1^\dagger c_2+c_2^\dagger c_1)\\
\hat{J}_x   & = &   \frac{1}{2}(c_2^\dagger c_2-c_1^\dagger c_1)\\ 
\hat{J}_y  & = &   \frac{i}{2}(c_2^\dagger c_1-c_1^\dagger c_2)
\end{eqnarray}
and setting $c_1^\dagger c_1+ c_2^\dagger c_2= \hat{N} =N$ (as the total number
is conserved), the Hamiltonian becomes
\begin{equation}
\hat{H}_2= \hbar\Omega\hat{J}_z+2\hbar \kappa\hat{J}_x^2  .
\label{ham}
\end{equation}
Here we have neglected terms proportional to $N$ and $N^2$ since they merely
correspond to a shift in the energy scale.  The Casimir 
invariant is 
\begin{equation}
\hat{J}^2=\frac{\hat{N}}{2}\left (\frac{\hat{N}}{2}+1\right ).
\end{equation}
This is analogous to an angular momentum model with total angular 
momentum given by $j=N/2$. 

The angular momentum operators have a simple physical interpretation. 
The operator $\hat{J}_z$ corresponds to the particle occupation number 
difference between the single-particle energy eigenstates. For example 
the maximal weight eigenstate $|j,j\rangle_z$ corresponds to all the 
particles occupying the highest single particle energy eigenstate, 
$\psi_2(x)$. The operator $\hat{J}_x$ gives the particle number 
difference between the localised states of each well. 
In fact the x-component of the position operator in the field representation is 
\begin{equation}
\hat{x} = \frac{2 q_0}{N}\hat{J}_x.
\end{equation}
Thus the maximal and minimal weight eigenstates of $\hat{J}_x$ correspond 
to the localisation of all the particles in one well or the other. 

\section{Homodyne Detection Scheme}

Figure \ref{fig1} illustrates the system under investigation here.  One of the
wells of the double well system is placed in one arm of an optical cavity.
The cavity is
strongly driven by a coherent field at the cavity frequency. We assume that on
the time scale of
tunneling oscillations, the cavity is heavily damped. The cavity field thus
relaxes to the
steady state on a much faster time scale than the BEC dynamics. This enables us
to make an adiabatic
elimination of the cavity dynamics. The cavity field is assumed to be far off
resonance with
respect to a dipole transition in the atomic species.  The effect of the atoms
is then entirely
dispersive and shifts the phase of the cavity field by an amount proportional to
the number of
atoms in the cavity at any particular time. If the atomic number in the cavity
oscillates, so will
the phase shift. Thus any tunneling of the condensate will be manifest in a
modulated phase shift
of the optical field exiting the cavity. To detect this phase shift we consider
a homodyne
detection scheme. The light leaving the cavity  is combined with the reference
beam and allowed to fall on a photodetector, which records the photocurrent.  If
there is a
difference in atom number of the two condensates, then coherent tunneling can
occur and the
homodyne current will be modulated at the tunneling frequency.  

Assuming that the 
incoming light is detuned from any atomic resonance, the interaction
Hamiltonian density is
\begin{equation}
\hat{\cal{H}}_I = \hat{\Psi}^\dagger(\vec{x})[H_{cm} - \hbar \mu g(\vec{x}) a^\dagger
a] \hat{\Psi}(\vec{x})
\end{equation}
where $a$, $a^\dagger$ are the cavity field operators, $g(\vec{x})$ is the
intensity mode function and $\mu = \Omega_R^2/4  \Delta$, with Rabi
frequency $\Omega_R$ and optical detuning $\Delta$.  $\hat{\Psi}(\vec{x})$ and 
$\hat{\Psi}^\dagger(\vec{x})$  are the atomic many-body operators, and $H_{cm}$
describes the centre of mass motion.

Introducing the condensate field operators $c$, $c^\dagger$ and averaging over
the optical mode function gives the interaction energy:
\begin{eqnarray}
\hat{H}_I &=& -\hbar\chi a^\dagger a c_1^\dagger c_1  \nonumber \\
	  &=& -\hbar\frac{N}{2} \chi a^\dagger a - \hbar\chi a^\dagger a
	  \hat{J}_x
\end{eqnarray}
where $\chi$ is the interaction strength.  If the optical mode has a beam waist 
$w$, then the interaction strength can be written
\begin{equation}  
\chi = \frac{\sqrt{2}\mu}{\sqrt{2(r_0/w)^2 + 1}}.
\end{equation}
For $N=100$ atoms, $\chi > 10^{-3}$ should give detectable phase-shifts ($0.1$
radians), and
should be experimentally feasible.  For example, with $r_0$ as above,
$w=30\mu m$, $\Delta/2\pi=100 MHz$, saturation intensity $I_s = 17 W/m^2$, optical
frequency $\omega/2\pi = 3.8 \times 10^{14} Hz$, atomic linewidth 
$\gamma_a/2\pi = 10^7 Hz$ and 
incident power $P=6 mW$, in a cavity $10cm$ long, then $\chi \approx 10^{-2}$.  
Larger values of $\chi$ may then be achieved by reducing the detuning or the 
incident intensity. 

The cavity is assumed to be driven by a strong coherent field of strength
$\epsilon$ and strongly damped at the rate $\gamma$.  Hence the master equation
for the whole system is, with $\hbar = 1$,
\begin{eqnarray}
\dot{\rho}_{tot} &=& -i\Omega[\hat{J}_z,\rho_{tot}]-i 2\kappa[\hat{J}_x^2,\rho_{tot}] \nonumber \\
	   &-& i(\delta - \frac{N\chi}{2})[a^\dagger a,\rho_{tot}] 
	   + i \chi [a^\dagger a \hat{J}_x, \rho_{tot}] \nonumber	\\
	   &-& i\epsilon[a^\dagger + a, \rho_{tot}] + \frac{\gamma}{2}(2a\rho_{tot}
	   a^\dagger - a^\dagger a \rho_{tot} - \rho_{tot} a^\dagger a)
\end{eqnarray}
where the initial cavity detuning $\delta = N\chi/2$ is chosen to remove the
$N$-dependent linear dispersion.  The optical field may now be adiabatically eliminated from the master equation 
\cite{MilJacWal94}:
\begin{eqnarray}
\dot{\rho} &=& -i\Omega[\hat{J}_z,\rho]-i 2\kappa[\hat{J}_x^2,\rho] \nonumber \\
	   &+& i \chi |\alpha|^2 [\hat{J}_x, \rho] - D[\hat{J}_x, [\hat{J}_x, 
\rho]]
\end{eqnarray}
where the coherent amplitude is $\alpha = -2i\epsilon/\gamma$ and 
$D = 8\chi^2\epsilon^2/\gamma^3$.  The double commutator represents a
decoherence produced by photon number fluctuations in the optical fields.  It is
a
quantum measurement back-action term consistent with the interpretation that the
optical field makes
a measurement on the condensate. In fact this last term destroys coherence in
the eigenbasis of
$\hat{J}_x$ and thus should inhibit tunneling oscillations. This is indeed true
for the ensemble
of measured systems described by the master equation. However, as we show below,
it is not true for
a particular realisation of a single measurement run.   The ensemble-averaged
effect of the
measurement can be seen in the operator moment equations (for $\kappa = 0$):
\begin{eqnarray}
<\dot{J}_x> &=& - \Omega <\hat{J}_y> \label{jx}\\
<\dot{J}_y> &=&  \chi |\alpha|^2 <\hat{J}_z> + \Omega <\hat{J}_x> - 
D<\hat{J}_y> \label{jy}\\
<\dot{J}_z> &=& - \chi |\alpha|^2 <\hat{J}_y> -D<\hat{J}_z>.\label{jz}
\end{eqnarray}
The terms with coefficient $\chi |\alpha|^2$ produce a precession around the
x-axis, which tends to inhibit coherent tunneling.  The effect of these terms
can be negated by adding a linear ramp, or tilt, to the double well potential.  
The $D$ terms cause a 
decay toward the origin, indicating decoherence, as expected.  If the system is
started in a number state with an equal number of atoms in each well, then no
tunneling will occur at all - these moments remain identically zero. 

When the wells are tilted, so that the precession around the x-axis is
suppressed, we can obtain equations for the second order moments:

\begin{eqnarray}
\dot{<\hat{J}_x^2>} &=& - \Omega <\hat{\Lambda}> \label{jxx}\\
\dot{<\hat{J}_y^2>} &=&   \Omega <\hat{\Lambda}> + 2D(<\hat{J}_z^2> - 
<\hat{J}_y^2>) \label{jyy}\\
\dot{<\hat{J}_z^2>} &=&  2D(<\hat{J}_y^2> - <\hat{J}_z^2>)\label{jzz}\\
\dot{<\hat{\Lambda}>} &=& 2\Omega (<\hat{J}_x^2> - <\hat{J}_y^2>) 
- D <\hat{\Lambda}>.\label{jxy}
\end{eqnarray}
where $\hat{\Lambda} = \hat{J}_x \hat{J}_y + \hat{J}_y \hat{J}_x$.

Thus even when the system is started with an equal number of atoms in each well,
the unconditional evolution of $ <\hat{J}_x^2> $ and $<\hat{J}_y^2>$ exhibit
oscillations initially.  For long times, the amplitude of these oscillations
decays due to $D$ and the system approaches the fixed point $<\hat{J}_x^2> = 
<\hat{J}_y^2> = <\hat{J}_z^2>$.  From this we see that the condensate
has on average a definite initial phase (which is clearly seen in the
second order moments, but not the first order moments), which is determined by 
the initial state.  In our simulations the initial state was chosen to be a
eigenstate of $\hat{J}_x$, which is not the only state which gives an equal
number of atoms in each well.  Presumably in a real experiment, the initial
state would be 
\begin{eqnarray}
|\psi(\tau)> = e^{-i\tau\hat{J}_z} |j,0>_x
\end{eqnarray}
where $\tau$ is a random variable uniformly distributed on the interval
$[0,2\pi]$.  This then implies a random initial phase.

A common technique for dealing with master equations describing open systems in 
quantum optics is to numerically simulate stochastic realisations of quantum 
trajectories.  This method has already been used by
several authors investigating the effect of measurement on the relative
phase of BECs\cite{JacColWal96,RuoWal97,RuoColGraWal98}, 
but these differ from the approach
used here in that we monitor the homodyne detection current.  The
resultant stochastic process is a diffusive evolution rather than the jump 
processes which occur in the direct detection of atoms or individual photons. 
The quantum trajectory method is a very appropriate one 
for the situation considered in this paper. We have one two-part condensate
system continuously
monitored by the homodyne detection scheme. If there is any phase difference
between the two parts
of the condensate it will be established in each run of the experiment. The
quantum trajectory
method enables us to simulate each run of an experiment. The master equation
however corresponds
to an average over many runs of the experiment and many homodyne current
records. For this reason
moments calculated directly from the master equation will show no evidence of
quantum tunneling if
there is no initial phase difference between the condensates. In contrast, as we
show, a single
run of the experiment can establish a self-consistent phase difference even if
no phase difference
in present initially. Such a `measurement induced' phase difference is manifest
in a measurement
induced tunneling current.  

The {\em conditional} master equation (that is the evolution conditioned on the
measurement result) for the optical field undergoing homodyne
detection is\cite{WisMil93,Wiseman94}
\begin{equation}
\left(\frac{d\rho_c}{dt}\right)_{field} = \gamma {\cal D}[a]\rho_c + 
\sqrt{\gamma} \frac{dW(t)}{dt} {\cal H}[a] \rho_c
\end{equation}
where $dW(t)$ is the infinitesimal Weiner increment.  In this equation, $\rho_c$
is the density matrix that is conditioned on a particular realisation of the
homodyne current up to time $t$.  Wiseman's superoperators
are defined as 
\begin{eqnarray}
{\cal D}[a]\rho &=& a\rho a^\dagger - \frac{1}{2}(a^\dagger a \rho + \rho 
a^\dagger a) \\
{\cal H}[a]\rho &=& a\rho + \rho a^\dagger - tr(a\rho + \rho a^\dagger)\rho.
\end{eqnarray}

The stochastic Shr\"{o}dinger equation, which describes the conditional 
evolution of the system is
\begin{equation}
d|\tilde{\Psi}_c(t)> = dt[-iH - \frac{1}{2}\gamma a^\dagger a +
I(t)a]|\tilde{\Psi}_c(t)>
\end{equation}
where $H = \Omega \hat{J}_z + 2\kappa \hat{J}_x^2 + H_I$ and 
$|\tilde{\Psi}_c(t)>$ is the {\em unnormalised} ket
describing the conditional state of the system.  The measured 
current is 
$I(t) =  \gamma <a + a^\dagger>(t) + \sqrt{\gamma}\xi (t)$, where the 
stochastic term $\xi (t)$ has the correlations
\begin{eqnarray}
<\xi (t)> &=& 0 \\
<\xi (t),\xi (t')> &=& \delta(t-t').
\end{eqnarray}
 Thus we can see how the system
evolution is conditioned upon the measured current.

Adiabatic elimination of the optical field, using eq(17) gives
\begin{eqnarray}
d|\tilde{\Psi}_c(t)> &=& dt[-iH - \frac{8\chi^2\epsilon^2}{\gamma^3}\hat{J}_x^2
+I(t)\hat{J}_x]|\tilde{\Psi}_c(t)> 
\label{SSEun} \\
I(t) &=& \frac{32\chi^2\epsilon^2}{\gamma^3}<\hat{J}_x>_c + 
\frac{4\chi\epsilon}{\sqrt{\gamma^3}}\frac{dW}{dt}(t).
\label{current}
\end{eqnarray}
Hence the oscillations in the occupation number between the two wells can be
determined from the measured current.

It is helpful to consider the
Shr\"{o}dinger equation for the {\em normalised} ket, which does not explicitly 
mention the detection current:
\begin{eqnarray}
d|\Psi_c(t)> &=& [-iHdt - \frac{8\chi^2\epsilon^2}{\gamma^3}(\hat{J}_x -
<J_x>_c)^2dt + \frac{4\chi\epsilon}{\sqrt{\gamma^3}}(\hat{J}_x - <J_x>_c)dW]
|\Psi_c(t)> \label{SSE}
\end{eqnarray}

The terms in the equation due to the measurement depend on the quantity 
$\hat{J}_x - <J_x>_c$.  This is minimal in semi-classical type trajectories for 
which $<\hat{J}_x^2>_c$ factorizes to $<\hat{J}_x>_c^2$.  Thus it may be 
expected that for some range of values of $\chi$, the stochastic measurement 
terms would drive
the system towards an oscillating trajectory for which $<\hat{J}_x^2>_c \simeq 
<\hat{J}_x>_c^2$.

\section{Simulations}

The results of the simulations are shown in figures 2 to 5.  Time is plotted
along the x-axis in dimensionless units (normalised by the inverse of the
tunneling frequency $\Omega$).  The strengths of the atom-atom collisions and 
the atom-field interaction were controlled by varying the normalised variables
$\bar{\kappa} = \kappa/\Omega$ and $\bar{\chi} =
\chi\epsilon/\sqrt{\Omega\gamma^3}$.  The parameters stated previously
give the range of measurement strengths used in the simulations ($10^{-4}<
\bar{\chi} < 0.1$) when the power of the optical field is varied from $0.06 mW$
to $6 mW$.  The mass of the particles is taken to be 
$m = 1.5 \times 10^{-25} kg$.

The measured current gives the conditional
dynamics of the the system.  However, in the conditional results shown, we plot 
$<\hat{J}_x>_c$, which is proportional to the current without the noise (eq
\ref{current}).  For clarity, the other moments, namely $<\hat{J}_y>_c$ and 
$<\hat{J}_z>_c$, are not plotted in the figures.  Except when $\bar{\kappa}$ 
is very
large, $<\hat{J}_y>_c$ follows $<\hat{J}_x>_c$, but with a $\pi/2$ phase
difference, and $<\hat{J}_z>_c$ remains at or close to zero.

The dynamics of the unmonitored system when started with all the condensate in 
one well has been discussed in previous work \cite{MilCorWriWal97}.  Basically 
when
there are no atom-atom interactions (ie $\bar{\kappa} = 0$), $<\hat{J}_x>$ oscillates
from $-N/2$ to $+N/2$.  When the interactions are present but only weak,
tunneling still occurs, but the amplitude quickly collapses due to nonlinear
dephasing.  The collapse is followed some time later by small revivals.  There
is a critical strength of collisions ($\bar{\kappa} N = 1$) at which the tunneling is
suppressed.  Above this value of $\bar{\kappa}$, the condensate is trapped in the well
in which it started, with only very small oscillations occuring in
$<\hat{J}_x>$.  

We expect to see similar behaviour in the current of the monitored system (eq 
\ref{SSE}). When 
$\bar{\kappa} = 0$, $<\hat{J}_x>_c$ 
oscillates as before, for weak atom-light coupling (ie $\bar{\chi} N \simeq < 
1$).  For stronger measurements, the resulting back action can be seen in the 
current.  For long times, the amplitude of the tunneling oscillations starts 
to fluctuate and a slow phase change is evident.  In the case when atom 
collisions are present, the effect of the measurements is to halt 
the collapse of the
oscillations seen in the unmonitored system.  The phase changes are also more
pronounced.  The effect of the critical value ($\bar{\kappa} N = 1$) is seen in the
suppression of the oscillations in the current above this value.  Figure 2 shows
the evolution of $<\hat{J}_x>_c$ for values of $\bar{\kappa}$ above and below the
critical value.
  
If the system is started with an equal number of atoms in each well, then we
expect no coherent tunneling in the absence of any detection apparatus. 
However, the presence of the field effects a measurement on the condensate
system.  This should establish a phase, which can be detected by measuring the
output current $I(t)$.  The simulations of eq(\ref{SSE}) show 
an oscillation in the current and,
for the optimum interaction strength, this can be established in a few
tunneling periods, for a small number of atoms.  The results for the case where
there are no atom collision, ie $\bar{\kappa} = 0$, are
shown in figure 3, for various measurement strengths $\bar{\chi}$.  The growth 
in oscillations occurs 
because, for large enough $\bar{\chi}$, $<\hat{J}_x>_c^2$ is driven to match 
$<\hat{J}_x^2>_c$, which
typically has large oscillations.  

If, as in figure 3(a) and (b),  the interaction strength is too small
($\bar{\chi} N \leq 0.1$), then generally the fluctuations are not large enough to
drive full tunneling and the current suffers small, rather irregular 
oscillations.  However even for large $\bar{\chi}$, when the oscillations in the
current are established, such as in figure 3(c), they are not guaranteed to stay
large in amplitude. 
This is because $<\hat{J}_z^2>_c$ undergoes what appears to be a random walk
which, because of the Casimir invariant, directly affects the amplitude of the
oscillations in $<\hat{J}_x^2>_c$ and $<\hat{J}_y^2>_c$. Consequently, since 
the 
measurement locks $<\hat{J}_x>_c^2$ onto the orbit of $<\hat{J}_x^2>_c$, this
changes the amplitude of the oscillations in the current.  Because of the 
random nature of the orbit of $<\hat{J}_z^2>_c$, the tunneling oscillations in 
the current over a certain time frame in a ``good'' run may be large, but in 
another with the same parameters the oscillations may be small and irregular in
amplitude.  

When the measurement is quite strong ($\bar{\chi} N \simeq 10$), as in figure 
3(d), the tunneling oscillations appear to be quite irregular.  However, a fourier
transform of the current picks out the tunneling frequency $\Omega$ very
strongly, so the fluctuations are mainly in the amplitude of oscillations, not
so much in the phase.  When the measurement strength is very large ($\bar{\chi} 
N = 40$), $<\hat{J}_x>_c$ 
still indicates a tunneling from one well to another, but the oscillations are 
no longer harmonic and appear quite random, both in frequency and amplitude.

The equations for the {\em unconditional} dynamics (eqs
\ref{jx}-\ref{jxy}) show a decay in the oscillations 
of $<\hat{J}_x^2>$ and $<\hat{J}_y^2>$ for long times, which increases with the 
measurement strength $\bar{\chi}$.  Figure 4 shows the unconditional dynamics 
for two different measurement strengths.  No such decay is seen in the 
individual
trajectories, but rather there is a diffusion in the phase of the oscillation
over long times which accounts for the decay in the mean evolution.  This change
in phase appears to be most rapid over the periods when the oscillations are
small in amplitude and most likely to suffer random fluctuations.

The presence of atom-atom interactions increases the phase diffusion, even for
quite weak collisions ($\bar{\kappa} N \leq 0.1$).  Figure 5 shows the evolution of 
$<\hat{J}_x>_c$ for various atom-atom interaction strengths $\bar{\kappa}$, above and
below the critical value.  The amplitude is also more 
irregular, and the fourier transform of the oscillations no longer shows a clear
peak at the expected tunneling frequency, but a group of random spikes centered
on the the tunneling frequency.  In figure 5(b), $\bar{\kappa} = 0.005$ which 
is close
to the value of $\bar{\kappa}$ calculated from the parameters given in previous 
sections.  Above the critical value of $\bar{\kappa} N = 1$, 
$<\hat{J}_x>_c$ suffers small, very irregular oscillations around the origin
(figure 5(c)).  This is quite different to the behaviour of $<\hat{J}_x>_c$ 
when the condensate was initially placed entirely in one well (figure 2(b)), 
in which case the condensated was trapped
in the well it started in and the critical value of ${\bar{\kappa}}$ marked quite a
sharp boundary (or bifurcation) between two different types of behaviour.  In
this case where the condensate is distributed equally between the wells, the
condensate is trapped in neither well, but remains across both, and the change 
in behaviour as the critical point is crossed is more continuous.  As 
$\bar{\kappa}$
increases past the critical point, we see a decrease in the overall amplitude of
the tunneling and in its regularity.  However, the critical value of
$\bar{\kappa}$
is still quite meaningful in this case an indication of when the strength of the
atom collisions significantly suppresses the tunneling.

\section{Concluding Remarks}

We have shown that this homodyne detection scheme, for an appropriate choice of
measurement strength $\chi$, could well be suitable to detect the relative
phase, in the form of Josephson-like
tunneling, between two condensates in a double well potential.  The dynamics of
the measured current reflect the tunneling of the condensate as well as the 
self-trapping effect caused by atom collisions.  It also demonstrates quite
vividly how a measurement can establish a (relative) phase in a system which
initially exhibits no phase information.  

\acknowledgments{The authors would like to thank H. M. Wiseman for the useful
discussions}

\begin{figure}
\caption{Schematic representation of the homodyne detection scheme to monitor
the
tunneling between two spatially separated condensates. One part of the
condensate is contained
in an optical cavity. The light in the cavity is well detuned from the atomic
resonance. The output
light from the cavity is detected by balanced homodyne detection.}
\label{fig1}
\end{figure}

\begin{figure}
\caption{Evolution of $<J_x>_c$ in the monitored system, for 
$N=100$ atoms all initially in one well and $\bar{\chi} = 0.01$.  In (a),
$\bar{\kappa}=0.005$ and in (b), $\bar{\kappa}=0.02$.}
\label{fig2}
\end{figure}

\begin{figure}
\caption{Evolution of $<J_x>_c$ in the monitored system, for 
$N=100$ atoms and $\bar{\kappa} = 0$.  In (a),
$\bar{\chi}=0.0001$, in (b), $\bar{\chi}=0.001$, (c), $\bar{\chi}=0.01$, and in 
(d), $\bar{\chi}=0.1$.}
\label{fig3}
\end{figure}

\begin{figure}
\caption{Unconditional evolution of second order moments, for 
$N=100$
atoms and $\bar{\kappa} = 0$.  In (a),
$\bar{\chi}=0.01$, in (b), $\bar{\chi}=0.1$} 
\label{fig4}
\end{figure}

\begin{figure}
\caption{Evolution of $<J_x>_c$ in the monitored system, for 
$N=100$ atoms, and $\bar{\chi} = 0.001$.  In (a), 
$\bar{\kappa}=0.0001$, in (b), $\bar{\kappa}=0.005$, and in (c) 
$\bar{\kappa}=0.02$.}
\label{fig5}
\end{figure}

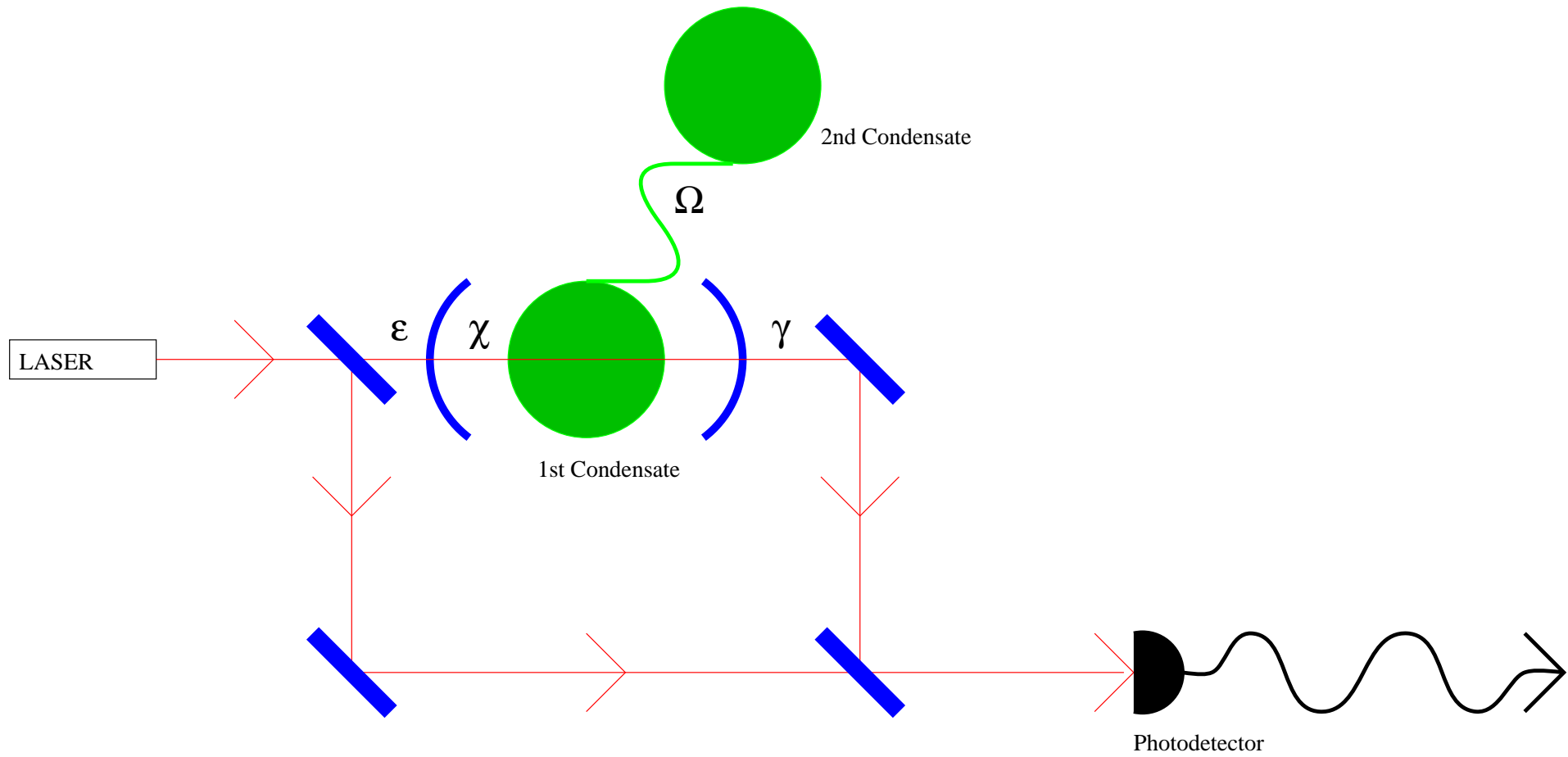
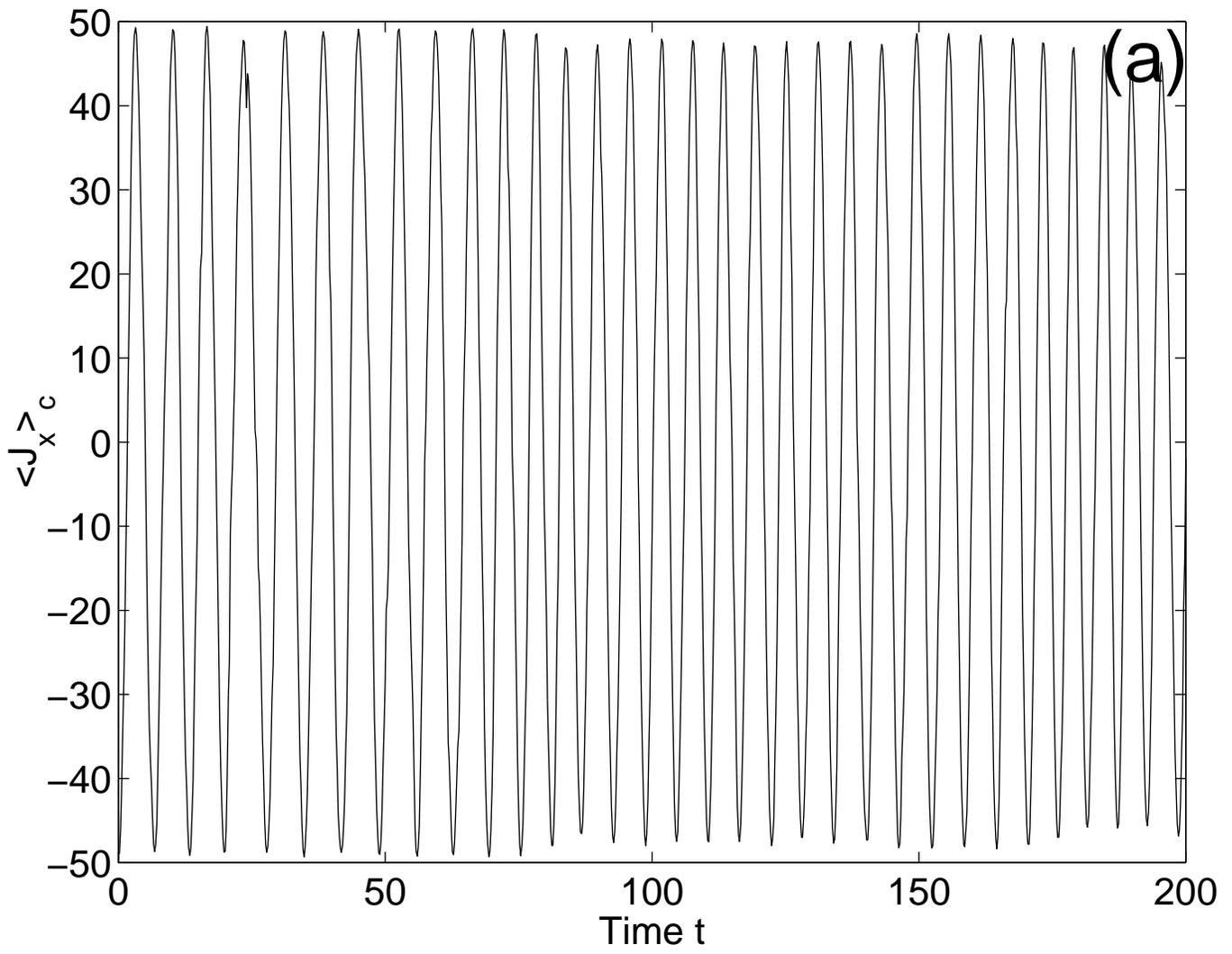
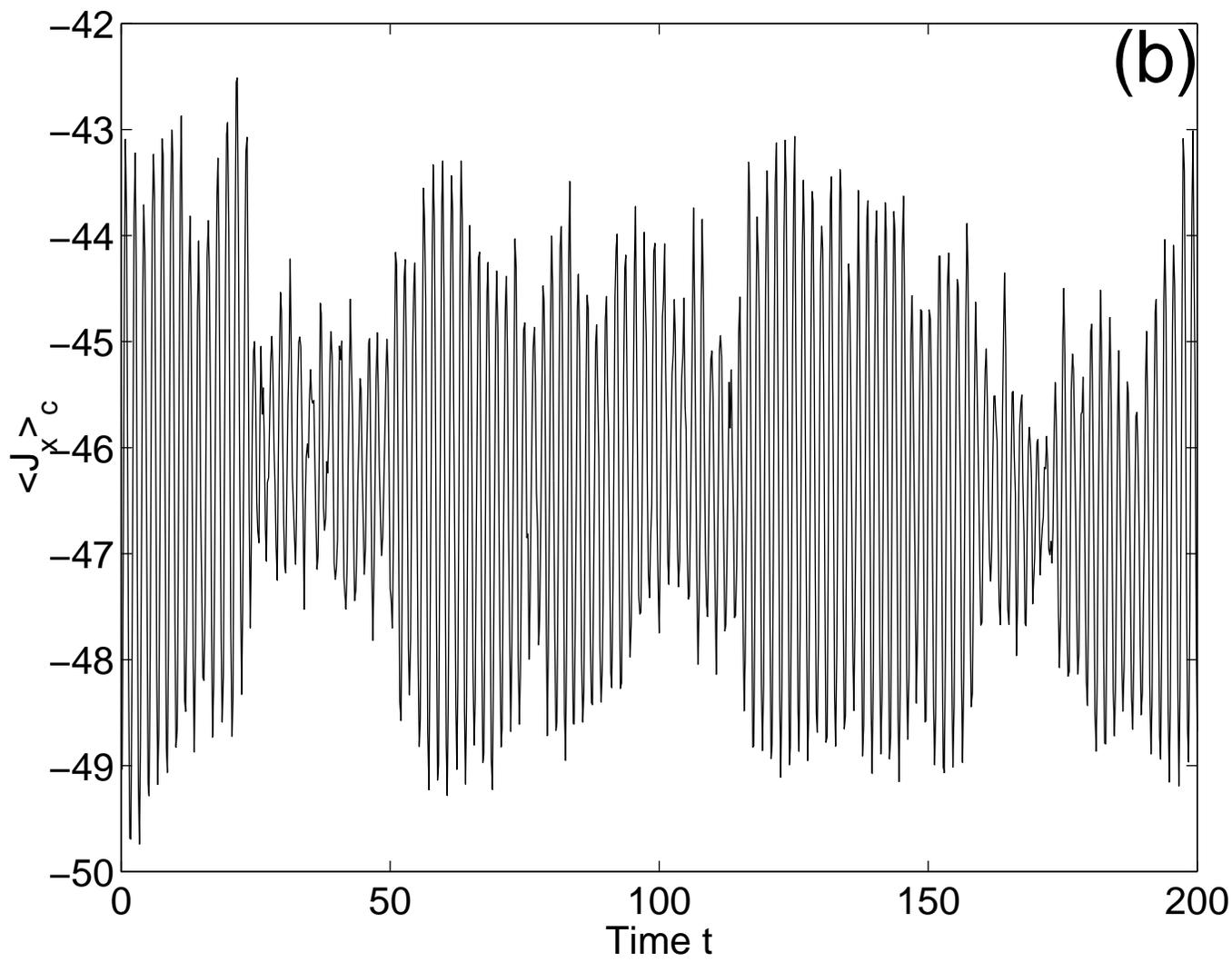
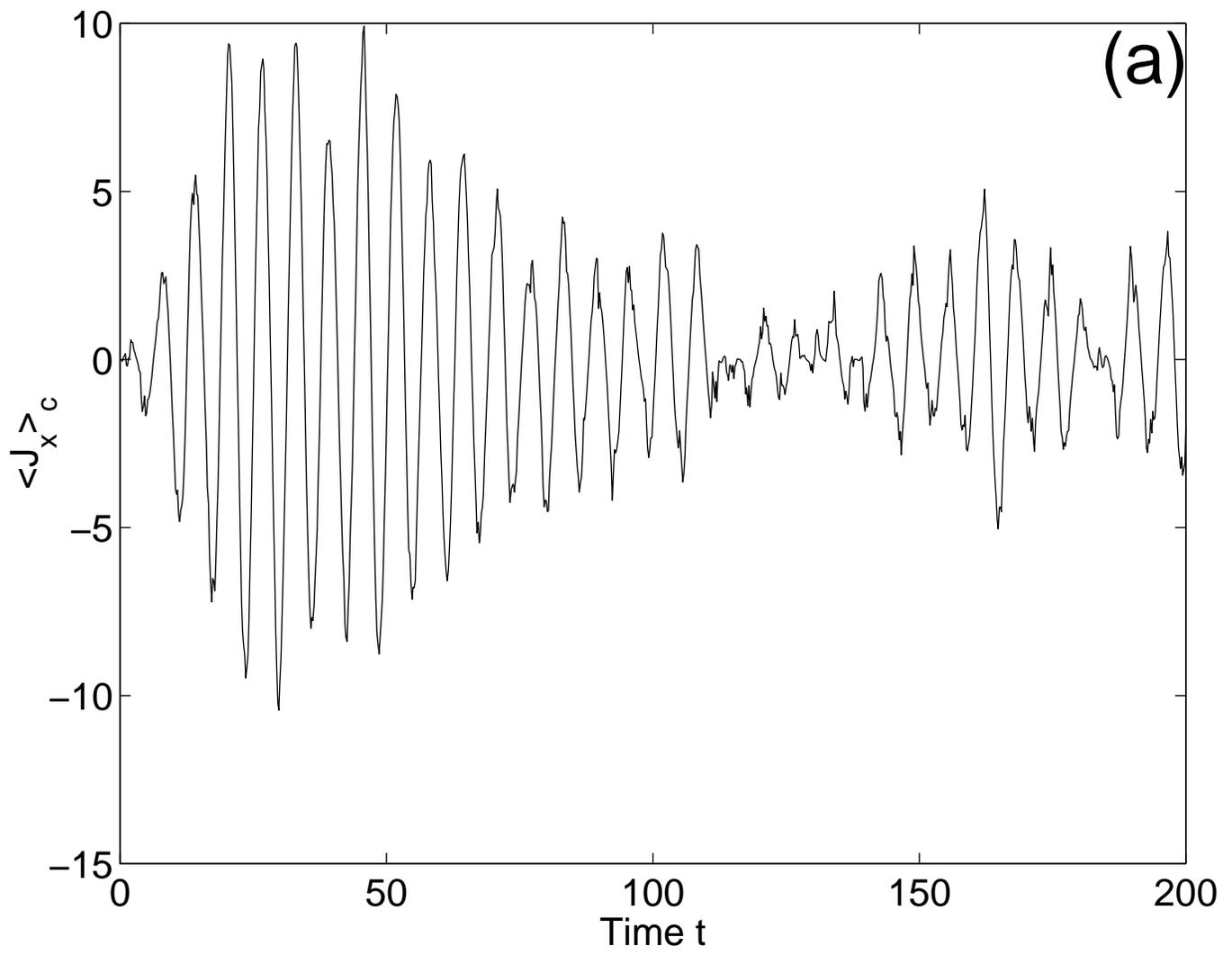
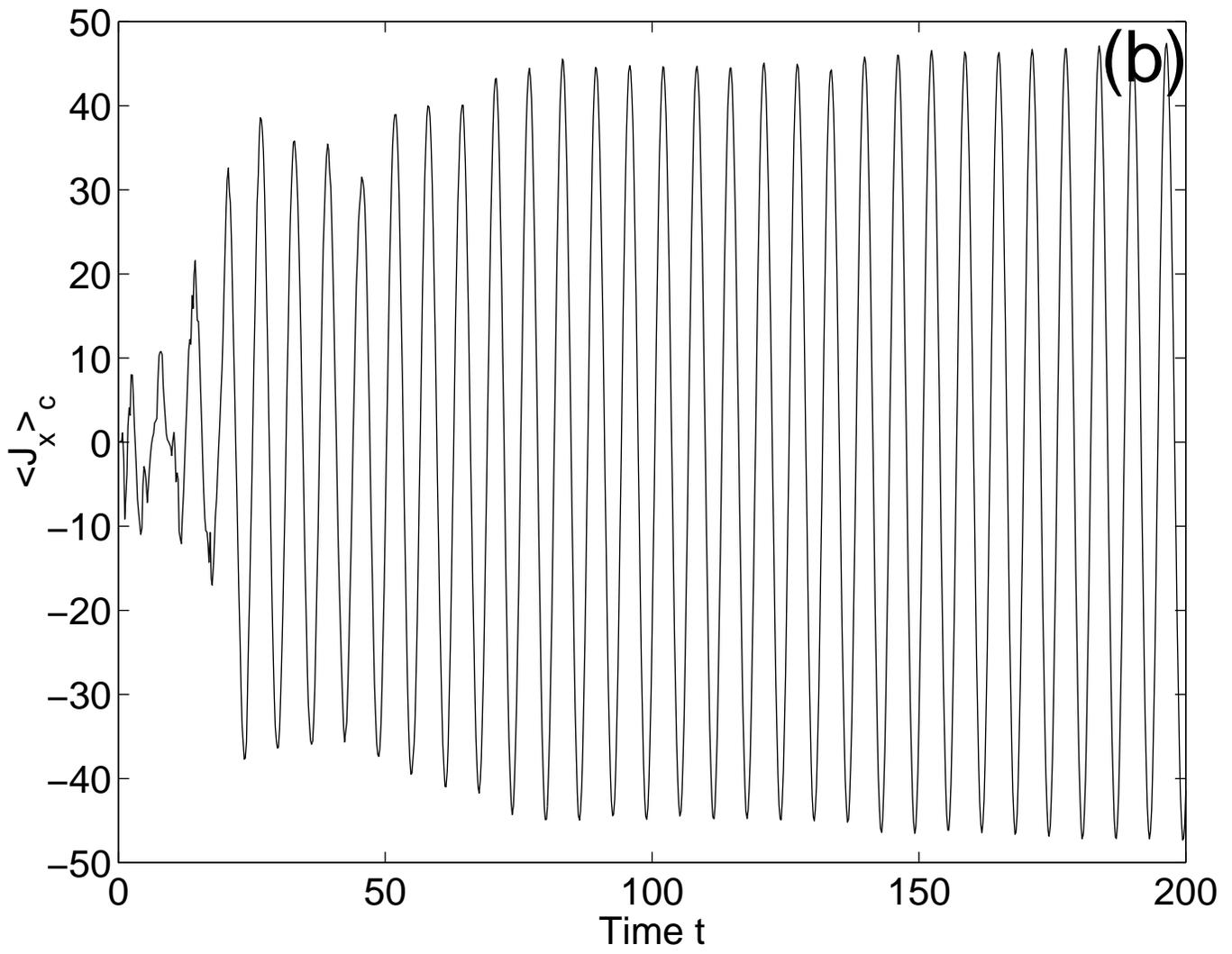
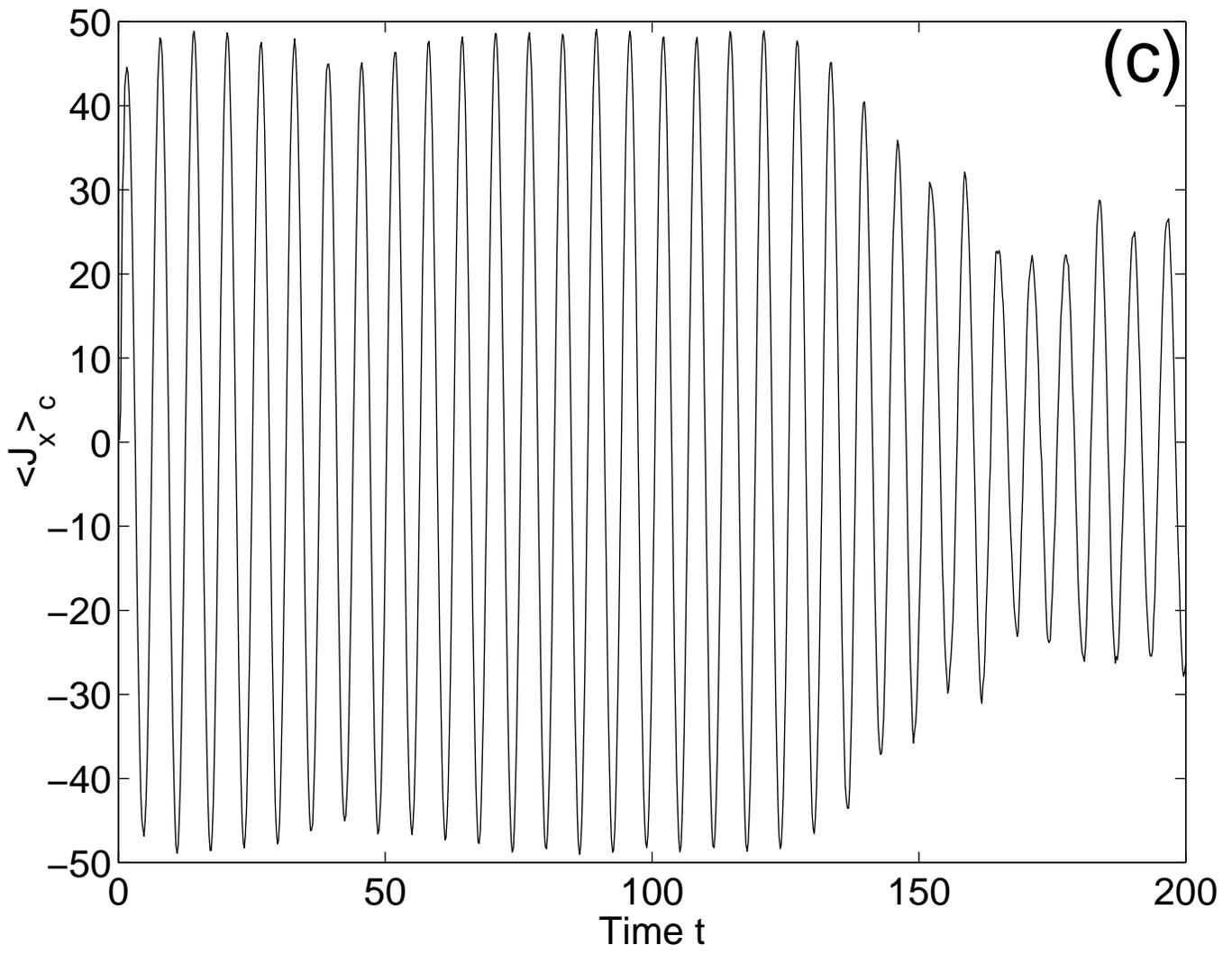
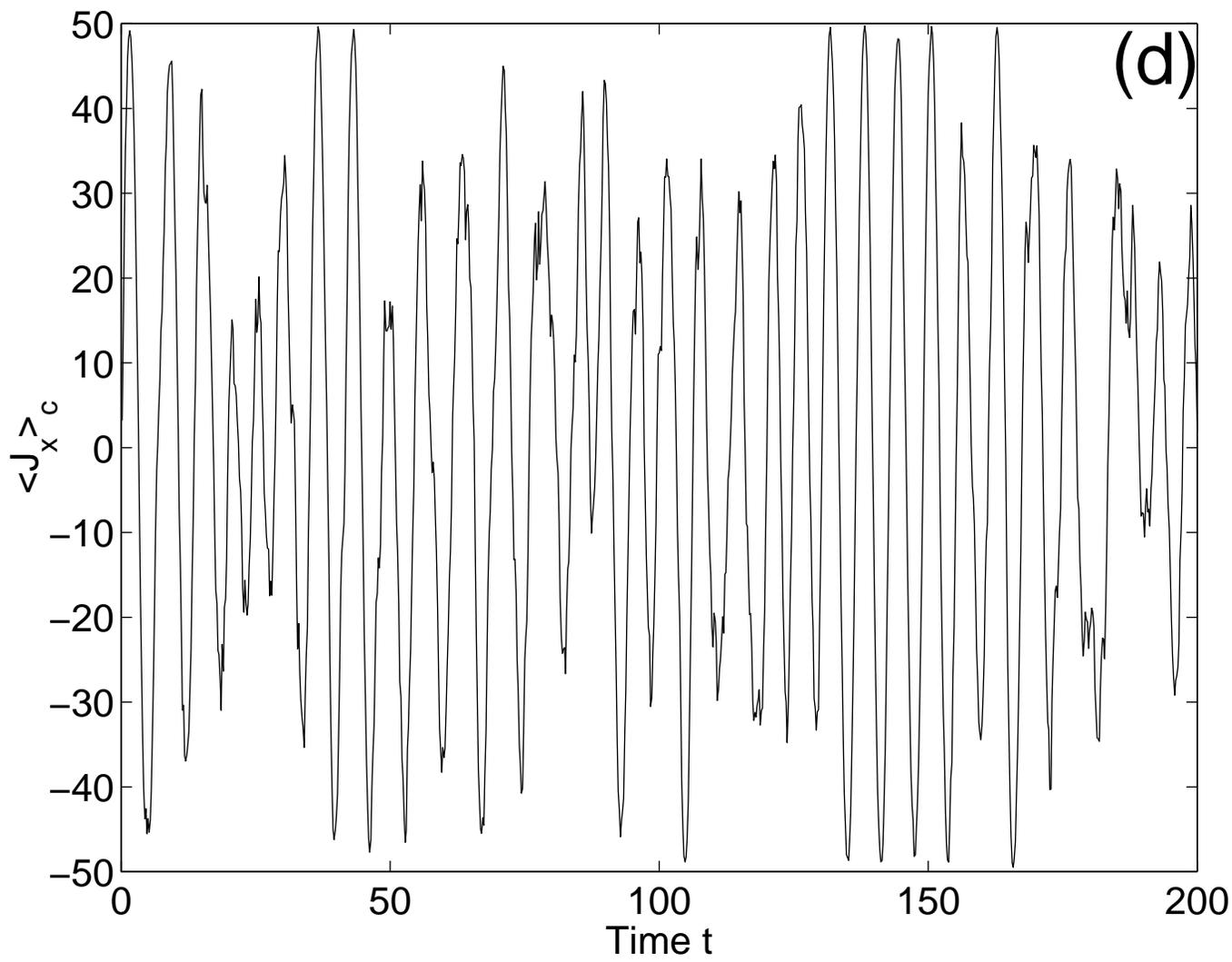
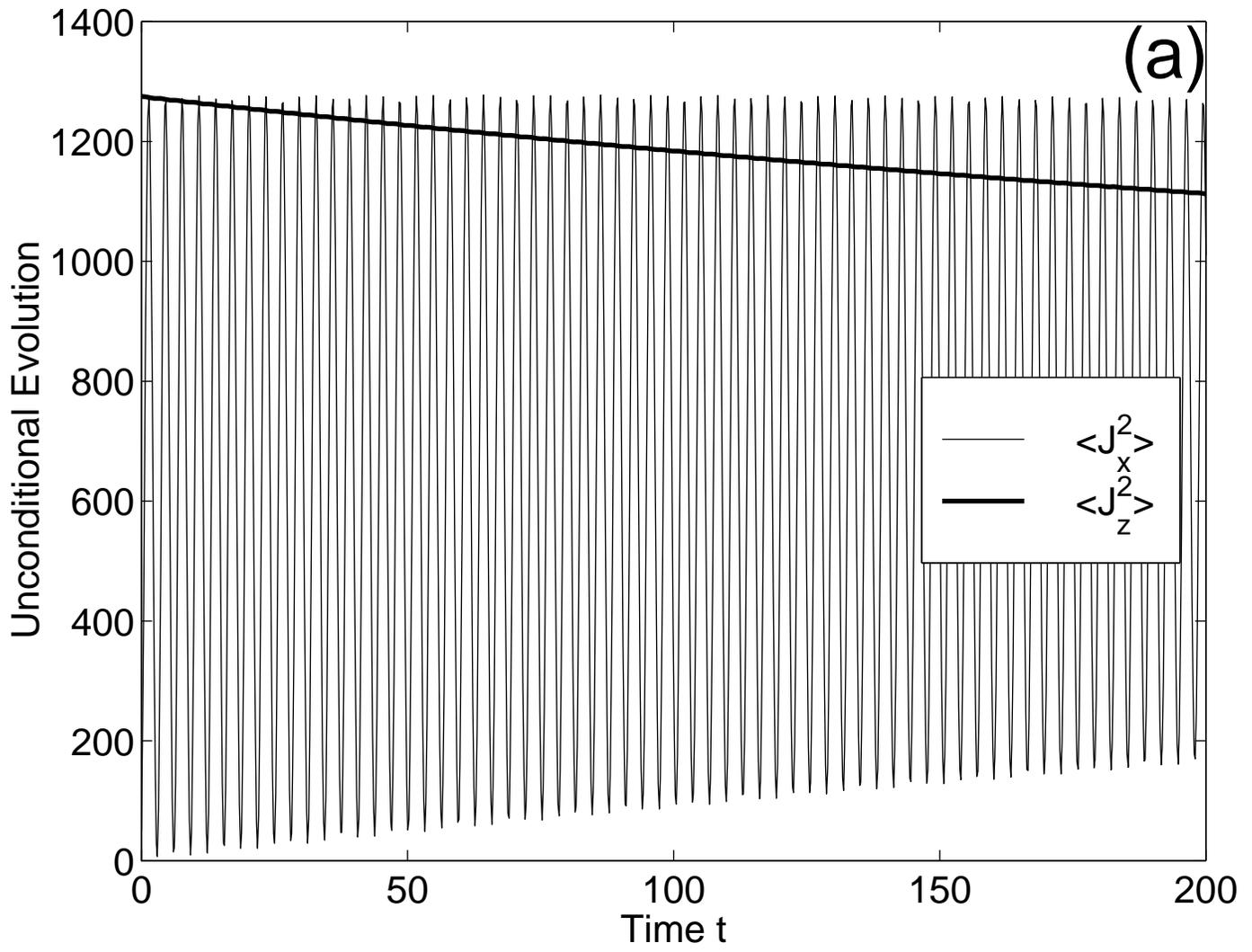
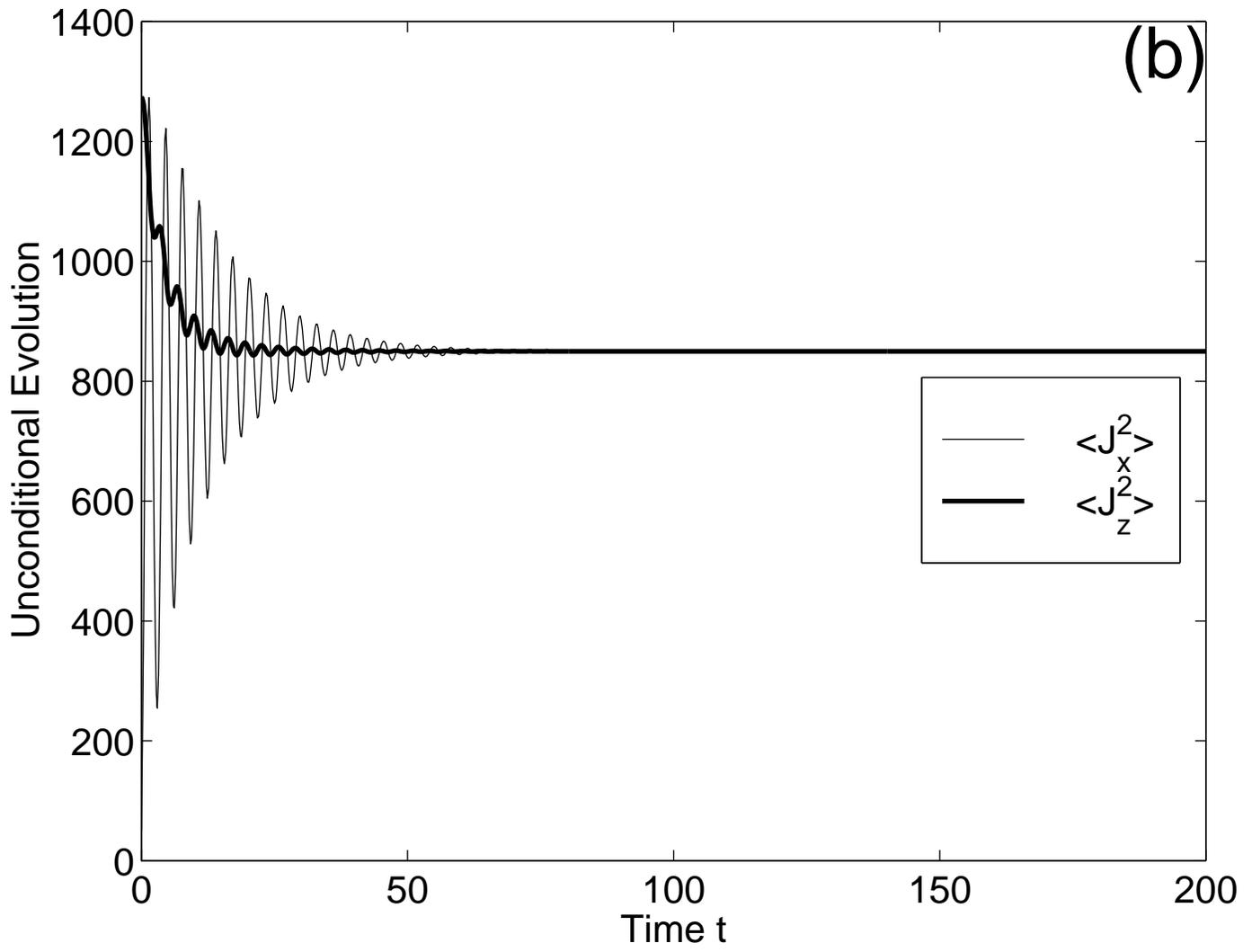
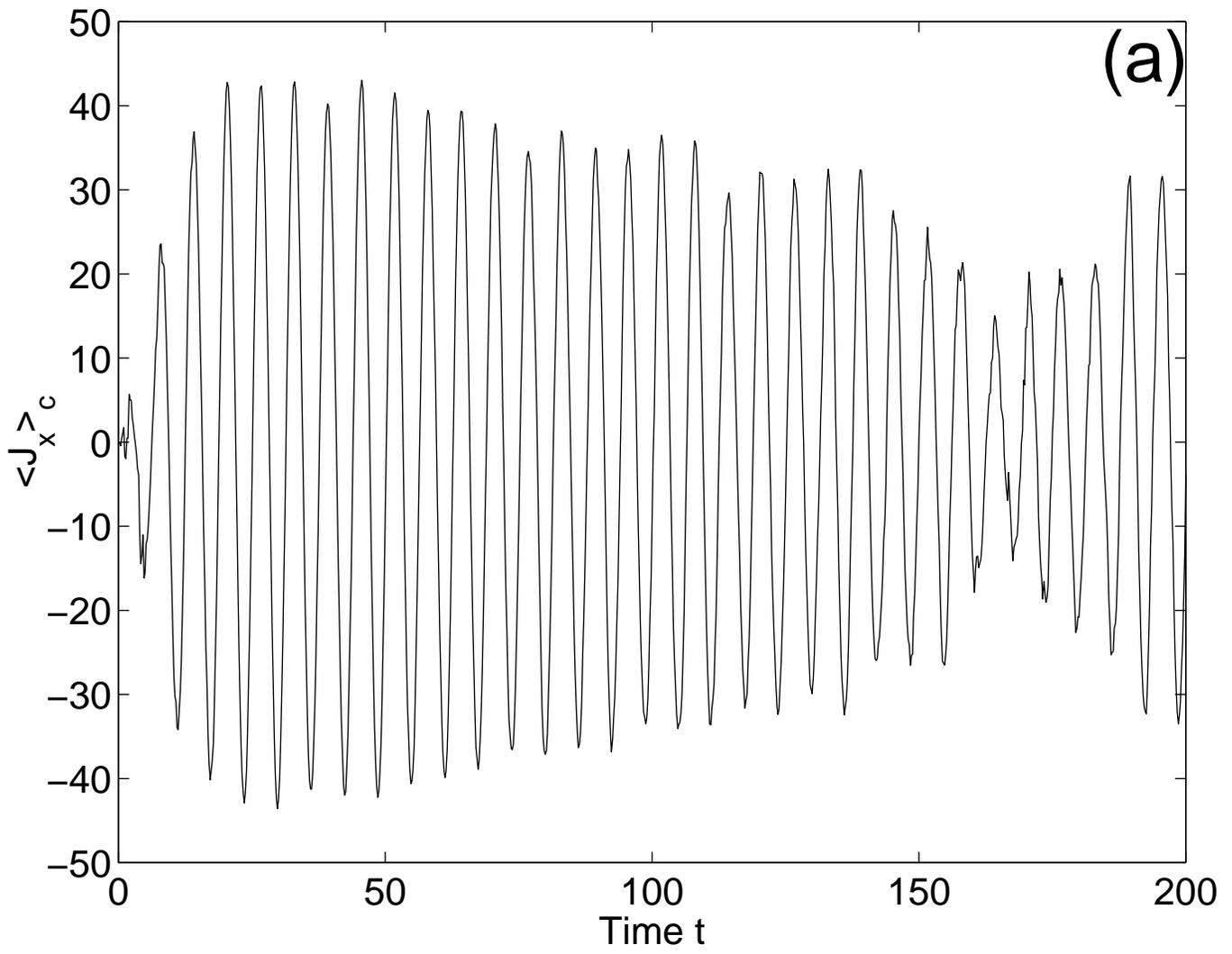
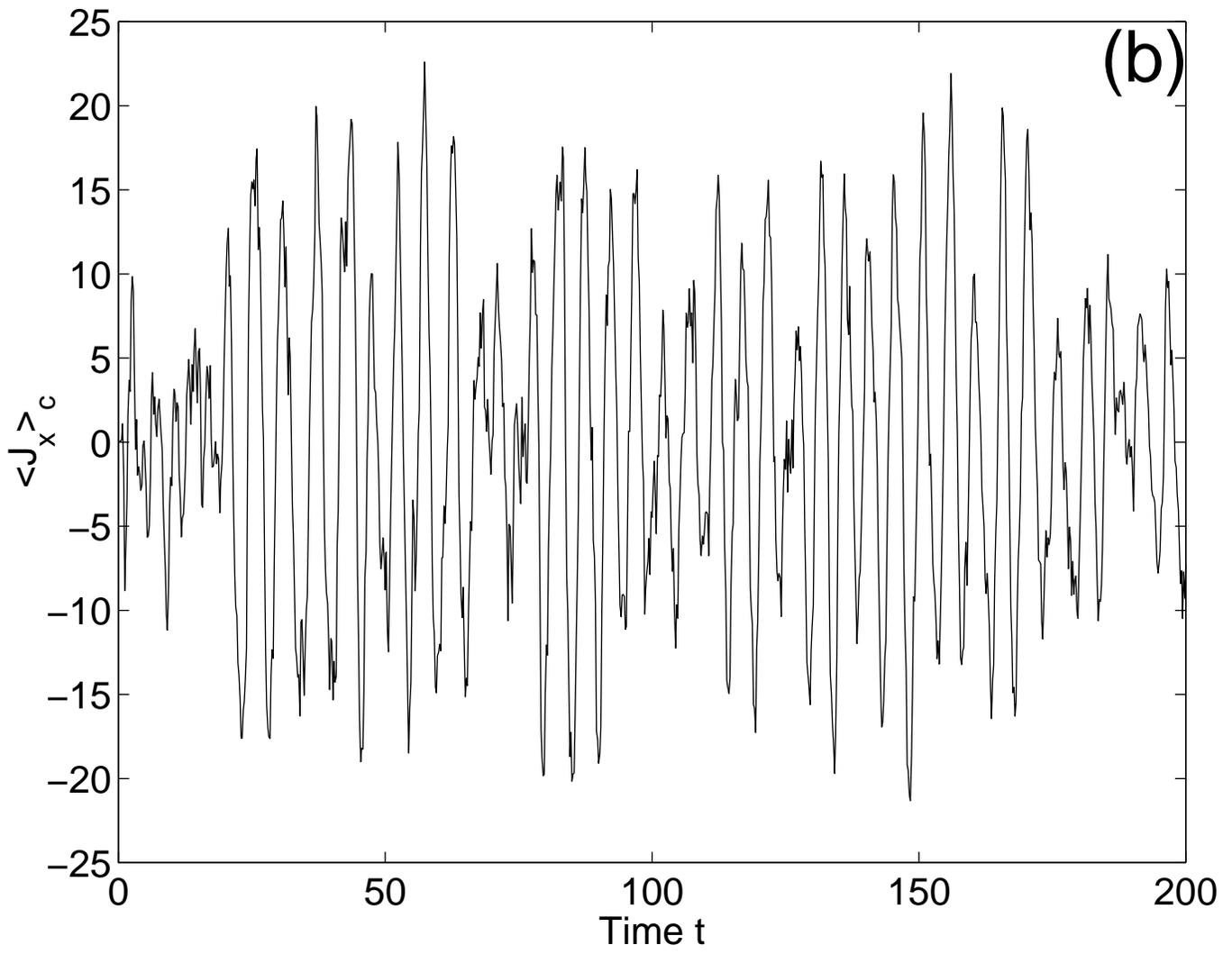
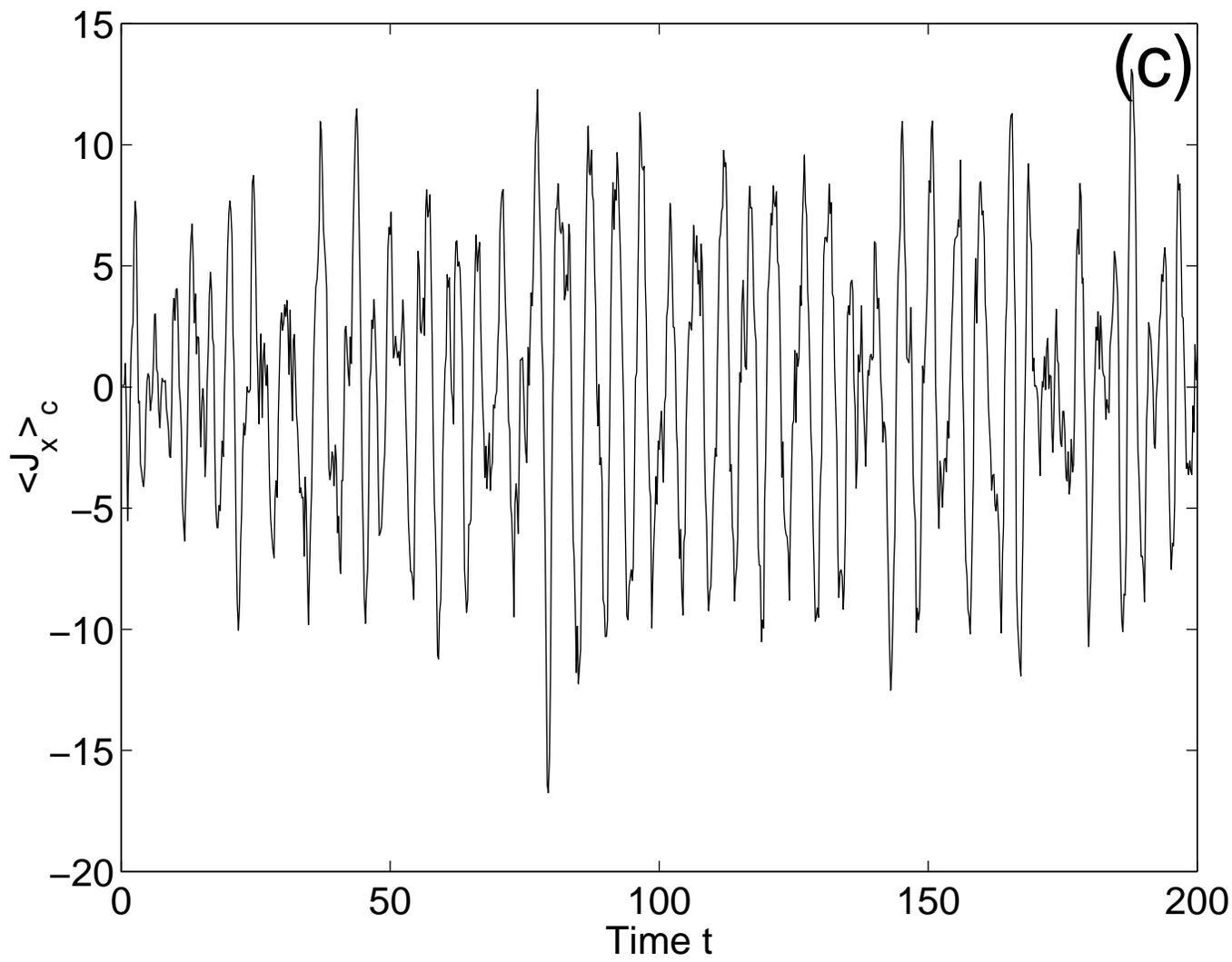
\end{document}